\documentclass[pdflatex,sn-mathphys-num]{sn-jnl}

\usepackage{graphicx}%
\usepackage{multirow}%
\usepackage{amsmath,amssymb,amsfonts}%
\usepackage{amsthm}%
\usepackage{mathrsfs}%
\usepackage[title]{appendix}%
\usepackage{xcolor}%
\usepackage{textcomp}%
\usepackage{manyfoot}%
\usepackage{booktabs}%
\usepackage{algorithm}%
\usepackage{algorithmicx}%
\usepackage{algpseudocode}%
\usepackage{listings}%

\theoremstyle{thmstyleone}%
\theoremstyle{thmstyletwo}%

\theoremstyle{thmstylethree}%

\begin{document}

\title[Article Title]{Towards transferable lightweight neuromorphic computing through a model-free temporal-switch framework}


\author[1,2,3]{\fnm{Zefeng} \sur{Zhang}}\email{21110850044@m.fudan.edu.cn}

\author[1,3]{\fnm{Chao} \sur{Li}}\email{lichao@ime.ac.cn}

\author[2,4]{\fnm{Siyao} \sur{Chen}}\email{21210180115@m.fudan.edu.cn}

\author[1,3]{\fnm{Pei} \sur{Chen}}\email{22112020077@m.fudan.edu.cn}

\author*[2,5,6]{\fnm{Bo-Wei} \sur{Qin}}\email{boweiqin@fudan.edu.cn}

\author*[1,3]{\fnm{Xumeng} \sur{Zhang}}\email{xumengzhang@fudan.edu.cn}

\author*[2,4,5,6]{\fnm{Wei} \sur{Lin}}\email{wlin@fudan.edu.cn}

\author*[1,3]{\fnm{Qi} \sur{Liu}}\email{qi\_liu@fudan.edu.cn}

\affil[1]{\orgdiv{State Key Laboratory of Integrated Chips and Systems}, \orgname{Fudan University}, \orgaddress{\city{Shanghai}, \postcode{200433}, \country{China}}}

\affil[2]{\orgdiv{Research Institute of Intelligent Complex Systems}, \orgname{Fudan University}, \orgaddress{\city{Shanghai}, \postcode{200433}, \country{China}}}

\affil[3]{\orgdiv{Frontier Institute of Chip and System}, \orgname{Fudan University}, \orgaddress{\city{Shanghai}, \postcode{200433}, \country{China}}}

\affil[4]{\orgdiv{School of Mathematical Sciences and Shanghai Center for Mathematical Sciences}, \orgaddress{\city{Shanghai}, \postcode{200433}, \country{China}}}

\affil[5]{\orgname{Shanghai Artificial Intelligence Laboratory}, \orgaddress{\city{Shanghai}, \postcode{200232}, \country{China}}}

\affil[6]{\orgdiv{State Key Laboratory of Medical Neurobiology and MOE Frontiers Center for Brain Science, Institute of Brain Science}, \orgname{Fudan University}, \orgaddress{\city{Shanghai}, \postcode{200032}, \country{China}}}


\abstract{

Lightweight neuromorphic computing offers a promising route to efficient AI, with particular benefits for resource-constrained edge deployments. However, its scalable deployment that can reliably transfer the expected performance has long been hindered by device-to-device variations, which necessitate costly and repeated re-training on new copies and undermine the practical advantages. To address this issue, we introduce a model-free temporal-switch (TS) framework to improve the direct transfer performance, without post-training calibration or adjustment. The TS framework provides a methodology to incorporate a broader spectrum of devices in the training process. In the validation using memristor-based reservoir computing, it enables high performance on unseen devices with a directly transferred readout. It achieves improved prediction in the representative Mackey--Glass benchmark, and the accuracy of 92.4\% in spoken digit classification. Its efficacy is validated across different memristor families and RC configurations. Theoretical analysis not only reveals the general computational mechanism underlying its efficacy, but also underlines its potential applicability to other physical platforms.
}


\maketitle

\section*{Introduction}\label{sec1}

Lightweight neuromorphic computing has attracted increasing interest in both academia and industry as a promising hardware-oriented route to efficient AI \cite{kudithipudi2025neuromorphic, kumar2020third, davies2018loihi}, particularly in resource-constrained edge scenarios where many distributed devices operate in parallel \cite{Li2024synsense, mansouri2021edge}. Among diverse lightweight neuromorphic computing paradigms, physical reservoir computing (RC) is particularly attractive \cite{yan2024emerging, liang2024physical}. It uses the intrinsic nonlinear and short-term memory dynamics of physical devices \cite{zhong2026physical, zhao2025smart, cui2025bioinspired, zhou2026protonic, jeon2025time}, such as dynamic memristors \cite{du2017reservoir, zhong2022memristor}, as computational black-boxes. By employing a simple three-layer architecture where only the readout is trained \cite{jaeger2004harnessing}, it enables useful computation based on measured states \cite{lin2025resistive, wang2023esgnn} while reducing the need for accurate device modelling and extensive parameter tuning \cite{sun2019understand, gao2021model}. Consequently, it offers an efficient and scalable way to harness the rich device dynamics for information processing.

\begin{figure}[h]
\centering
\includegraphics[width=1\textwidth]{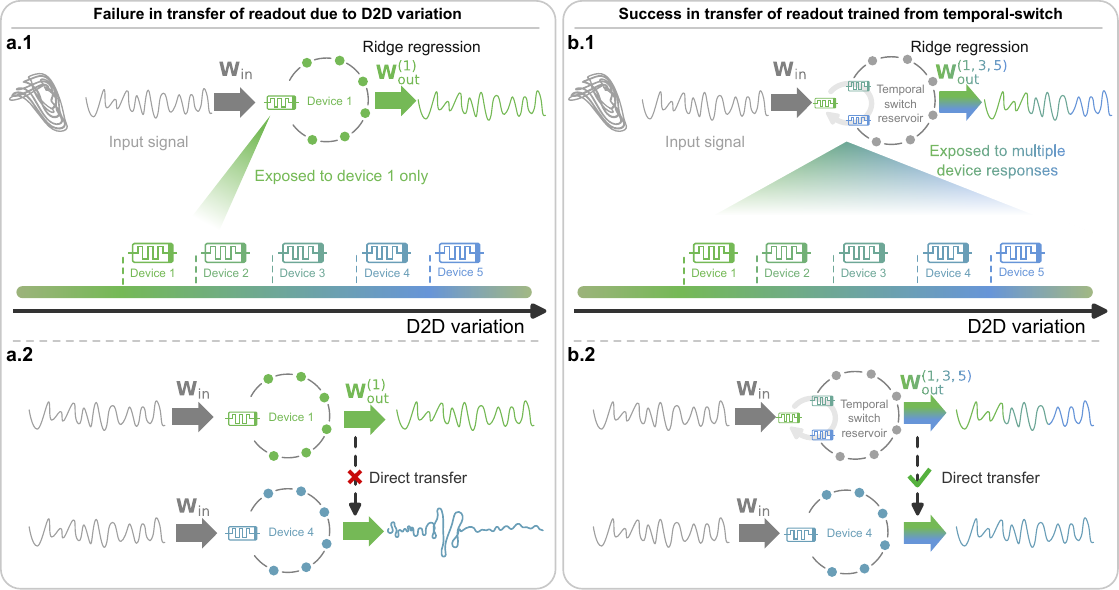}
\caption{Direct transfer of readout in the classical and the TS framework. ({\bf a}) The limited transferability of the readout in the classical framework. ({\bf a.1}) The classical training is only exposed to responses of one specific device (e.g., device 1); ({\bf a.2}) the nominally identical RC based on another unseen device (e.g., device 4) fails in generating the expected output with the transferred readout ${\bf W}_{\rm out}^{(1)}$. The superscript indicates the readout is trained on device 1 solely. ({\bf b}) The improved transferability by the TS framework. ({\bf b.1}) The TS framework exposes the training to responses of multiple different devices. ({\bf b.2}) The unseen device (e.g., device 4) can generate the expected output with the transferred readout ${\bf W}_{\rm out}^{(1,3,5)}$. }\label{fig1}
\end{figure}

However, for such systems to move beyond individually optimized demonstrations towards reliably deployable hardware, the trained readout should be transferable across nominally identical hardware instances rather than tied to one specific physical device \cite{moon2019temporal, zhong2021dynamic}. It is a central and persistent challenge because of the widespread and inevitable device-to-device (D2D) variation caused by fabrication or initialization stochasticity \cite{rajendran2014improving, garcia2014building, yakopcic2013generalized}. In the classical training framework, the training is performed on a single device (or a single device set, for a more complex configuration). In this case, the readout is optimized for the measured state generated by that particular reservoir, exposed to the dynamics of one device/device set only (Fig. \ref{fig1}a.1). However, the D2D variation can alter the transformation of external inputs to the response states among nominally identical physical reservoirs. It further leads to differences in the states, the state covariances and the state--target correlations seen by the readout during training. Consequently, a readout trained on one physical reservoir becomes device-specific and causes performance degradation when directly transferred to another nominally identical reservoir which is fabricated with the same design and operated under the same nominal conditions and settings  \cite{moon2019temporal, zhong2021dynamic} (Fig. \ref{fig1}a.2). If each physical RC requires its own retraining, calibration, or repeated parameter adjustment for expected performance, its practical advantage is substantially weakened. Its per-device adjusting requirement would make it difficult to use as a scalable, replaceable, and distributed hardware module. Hardware-aware training \cite{wright2022deep}, noise-injection methods (including the straight-through-estimator, STE) \cite{wan2022compute, joshi2020accurate}, and surrogate-gradient \cite{cramer2022surrogate} approaches have provided useful strategies in other neuromorphic computing systems when hardware non-idealities can be represented by explicit, differentiable, or statistically parametrized software models (`isomorphic' \cite{momeni2025training}). In physical reservoir computing, however, the reservoir is often used as a computational black-box (`broken-isomorphism' \cite{momeni2025training}), and the D2D variation appears directly in the measured reservoir states rather than as a simple perturbation of trained network weights. Therefore, it still calls urgently for an effective training framework that improves readout transferability directly from measured physical states, without requiring an explicit device model or retraining on test devices.

Here we introduce the temporal-switch (TS) framework for improving direct readout transfer under D2D variation. It provides a methodology to expose a broader spectrum of device dynamics to the trained readout (Fig. \ref{fig1}b.1), with motivation from the multi-source pooled training \cite{hu2020domain, wang2023generalizing}. Reservoir states are measured from multiple devices or device sets that span the expected variation range, and temporally concatenated to optimize one shared readout. This readout is then directly deployed on an unseen device or device set of the same dimension to generate the expected output (Fig. \ref{fig1}b.2). Building on pooled regression over measured states, the TS framework establishes multi-reservoir sampling as a model-free route to direct transfer with recovered performance. In this framework, D2D variation is converted from an uncontrolled deployment mismatch into sampled diversity, allowing one shared readout to learn from the state–target correlations and covariance structures generated by different physical reservoirs.

In this work, we validate the TS framework using the dynamic ${\rm TiO}_x$- and ${\rm NbO}_x$-based reservoirs \cite{zhong2021dynamic} (Figs. S1-S3, detailed in Methods) with quantified D2D variation. In the Mackey–-Glass (MG) prediction benchmark \cite{mackey1977oscillation} and spoken digit classification \cite{zhong2021dynamic}, two representative temporal prediction and classification tasks, the TS framework reduces degradation caused by the direct transfer in the classical framework. Meanwhile, it demonstrates more reliable performance in improving direct transfer compared with optimally regularized single device/device set transfer and other multi-device baselines, including the ensemble and state-average methods. Analyses on the number of training devices, performance recovery, and boundaries marked by D2D variation indicator further show that the improvement is strongest when deployment reservoirs fall within the  variation range represented during training. By shifting readout learning from each individual deployment device/device set to a combination of representative devices/device sets, the TS framework establishes a methodological basis for transferable and reusable readouts across physical reservoirs, moving physical RC closer towards scalable and application-oriented hardware deployment.

\section*{Results}\label{sec2}

\subsection*{Memristor characteristics and D2D variations}

\begin{figure}[h]
\centering
\includegraphics[width=1\textwidth]{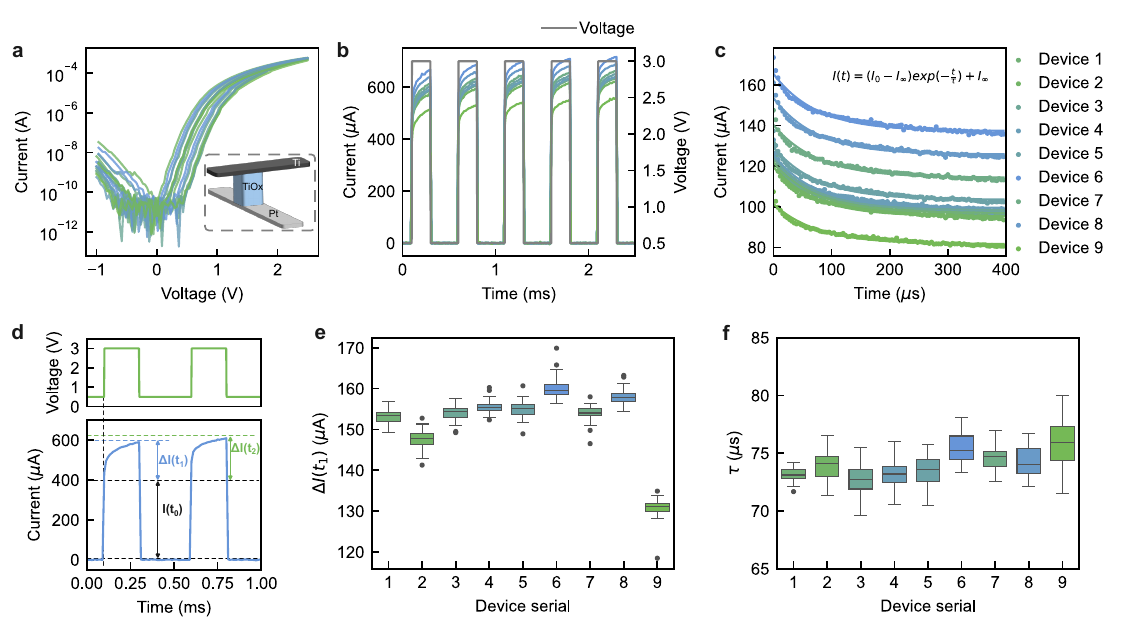}
\caption{Characteristics of ${\rm TiO}_x$ memristors. ({\bf a}) I--V characteristics of the randomly selected nine devices. In the inset dashed box is the device structure. Different curve colour corresponds to different device. ({\bf b}) Programming response under multiple pulses. The applied voltage pulses are shown in grey. ({\bf c}) Decay process. An exponential function is fitted to calculate the decay time constant $\tau$. ({\bf d}) Quantification of programming response through the increments of electrical current in the first pulse [$\Delta I(t_1)$] and the second pulse [$\Delta I(t_2)$]. ({\bf e}) Statistics of $\Delta I(t_1)$ of the nine devices (50 trials). ({\bf f}) Statistics of the fitted $\tau$ (20 trials).}\label{fig2}
\end{figure}

Two different families of dynamic memristors, the ${\rm TiO}_x$ \cite{li2023anticipative} and the ${\rm NbO}_x$ memristors, are fabricated for the following demonstration. They are all fabricated horizontally in identical crossbar structure (Fig.~S4) with the 5 $\mu$m $\times$ 5 $\mu$m overlapping area (detailed in Methods). The ${\rm TiO}_x$ and the ${\rm NbO}_x$ memristors are vertically assembled in sandwich structures of Ti/${\rm TiO}_x$/Pt and Au/${\rm NbO}_x$/Au, respectively (Figs.~\ref{fig2}a and S5a). The I--V curves characterize their intrinsic nonlinear physical dynamics (Figs.~\ref{fig2}a and S5d). Besides, the accumulating electrical currents during the pulse programming (Figs.~\ref{fig2}b and S5e), and the descending ones in the decay process (Figs.~\ref{fig2}c and S5b), characterize the devices' dynamic responses to external stimulus and short-term memories (detailed in Methods). The two are key requirements to build a physical RC~\cite{zhong2021dynamic, appeltant2011information}.

At the same time, the randomly collected devices exhibit obvious D2D variations within each family. While rapidly indicated by the distinguishably different I--V curves~\cite{rajendran2014improving, garcia2014building, yakopcic2013generalized}, the D2D variations are further characterized by the quantified details in pulse programming and decay process. On one hand, under the same voltage pulse, the increment of electrical current corresponds to the rate of conductance change (Fig.~\ref{fig2}d); on the other hand, the rate of decay (or decay time constant) can be calculated by the commonly used exponential regression (Fig.~\ref{fig2}c). Here we denote the increments of electrical currents in the first and the second pulses as $\Delta I(t_1)$ and $\Delta I (t_2)$, respectively, and the decay time constant as $\tau$. The analyses of variances (ANOVA) show statistical significance that the devices from the same family (${\rm TiO}_x$ or ${\rm NbO}_x$) do not share the same $\Delta I(t_1)$ (Figs.~\ref{fig2}e and S5f; ${\rm TiO}_x$: $p$-value=$1.3\times{10}^{-275}$; ${\rm NbO}_x$: $p$-value=$8.6\times{10}^{-72}$) or $\Delta I(t_2)$ (Fig.~S7; ${\rm TiO}_x$: $p$-value=$6\times{10}^{-283}$; ${\rm NbO}_x$: $p$-value=$1.1\times{10}^{-71}$). In other words, there are evident D2D variations in the programming responses. Furthermore, the intra-class correlation coefficient (ICC) analysis further reveals that D2D variations in decay time are relatively weak compared with the cycle-to-cycle (C2C) variations (Figs.~\ref{fig2}f and S5c; ${\rm TiO}_x$: ICC=0.311; ${\rm NbO}_x$: ICC=0.091), whereas those in the programming response are clear and pronounced (${\rm TiO}_x$: ICC=0.952; ${\rm NbO}_x$: ICC=0.990), dominating the overall D2D variations. To briefly summarize, the ${\rm TiO}_x$ and ${\rm NbO}_x$ memristors demonstrate the key properties required to implement physical RC, while exhibiting dominant D2D variations in their programming responses (statistical analyses detailed in Methods).

\subsection*{Direct-transfer improvement in temporal prediction}

\begin{figure}[h]
\centering
\includegraphics[width=1\textwidth]{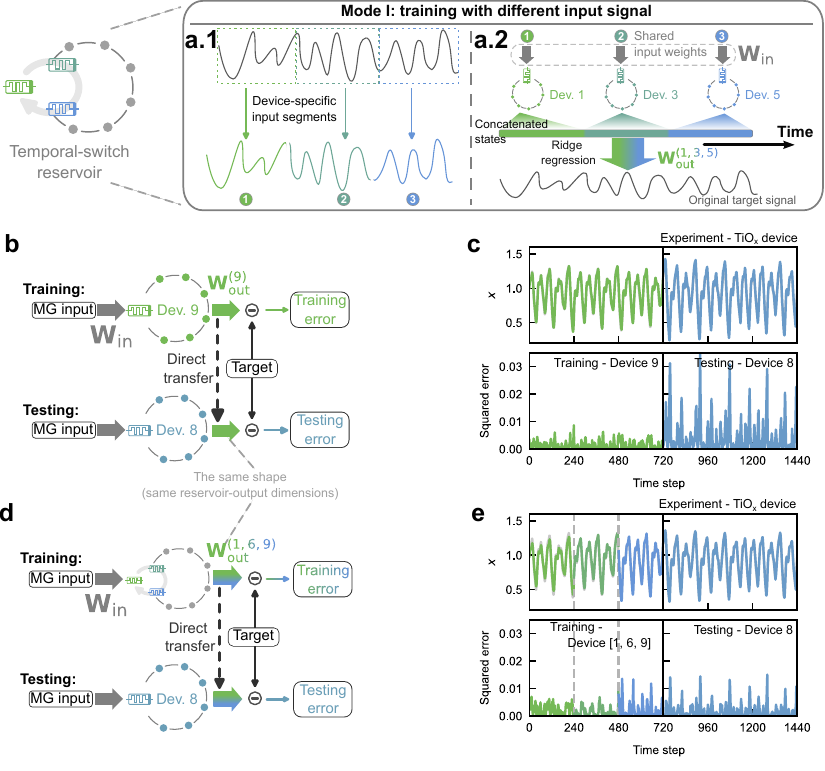}
\caption{Temporal-switch in the Mackey--Glass benchmark. ({\bf a}) The working mode I of the TS training. ({\bf a.1}) The original input signal is divided into device-specific segments; ({\bf a.2}) the readout is trained with the concatenated states from different devices as responses to the corresponding input. ({\bf b}) The classical training and transfer of the readout. ({\bf c}) The prediction and error by the RC of device 8 (testing NRMSE=0.270). ({\bf d}) The training and transfer process in the TS framework. ({\bf e}) The prediction and error by the RC of device 8 in the TS framework (testing NRMSE=0.174).}\label{fig3}
\end{figure}

We first evaluate the TS framework in the Mackey--Glass (MG) prediction benchmark \cite{hu2023distinguishing, mackey1977oscillation}, where the transferability of a trained readout can be directly quantified by the prediction error on unseen devices. Figure \ref{fig3}a shows a working procedure of the TS framework in the benchmark. The original input signal is split into $n$ segments of the same length (Fig.~\ref{fig3}a.1), and each segment is applied to one selected device using the same time-multiplexing \cite{appeltant2011information} (see Methods). The corresponding reservoir states are measured from the devices and concatenated sequentially along the temporal axis, to form a multi-device state matrix. A single readout is then trained by ridge regression using these temporally concatenated states and the original target signal (Fig.~\ref{fig3}a.2). Because the TS procedure only operates on the measured reservoir states and does not require an explicit device model, it remains model-free and naturally compatible with physical RC, where devices are used as black boxes to generate processed data. In addition, it keeps the total input length, the number of reservoir-state samples, and the sizes of the reservoir and readout identical to those in the classical framework, thereby isolating the contribution of multi-device response sampling rather than performance gains from increased training data or network scale. In the inference/testing phase, the readout trained from multi-device responses is directly transferred to an unseen device, without retraining or any parameter adjustment on test device (Fig.~\ref{fig3}d). The procedure is almost the same as that of the classical framework whose only distinction is the readout trained using states from a single device (Fig.~\ref{fig3}b). The complete workflow of the TS framework is summarized in Fig.~S8 and detailed in Methods, highlighting that temporal-switch operates in the training phase, whereas inference is performed on an unseen device/device set without increasing the reservoir size.

The evaluation is implemented experimentally by comparing the prediction precision of the same unseen ${\rm TiO}_x$-based RC (e.g., device 8) between the two frameworks. In the classical framework, the transferred readout trained on a single device [e.g., ${\bf W}_{\rm out}^{(9)}$ trained with device 9] produces a clear mismatch between the predicted and target MG signals. It results in large squared error and testing NRMSE of 0.270 (Fig.~\ref{fig3}c). The performance degradation is not restricted to this selected pair, but appears systematically across device pairs with clear D2D variation (Fig.~S9a). By contrast, when the readout is trained using the TS framework [e.g., ${\bf W}_{\rm out}^{(1,6,9)}$] with responses collected from devices 1, 6 and 9, the same unseen device produces a substantially improved precision, with the testing NRMSE of 0.174 (Fig.~\ref{fig3}e).

Furthermore, the transfer improvement achieved by the TS framework is not specific to the single-channel RC configuration alone, but also verified for the multi-channel parallel RC \cite{zhong2021dynamic}. In the multi-channel configuration, the readout is trained on the spatially joint states from multiple channels, and the channel-wise device combination defines an ordered device set (see Methods). Due to D2D variations, different device sets generate different joint state representations in response to the same input. Consequently, a readout trained in the classical framework on one device set incurs substantial performance degradation when directly transferred to an unseen device set (Fig. S9b). In the example using the three-channel RC (Fig. S10a), the transferred readout trained on one device set produces a prediction that clearly deviates from the target on another device set (Fig. S10b), with the testing NRMSE of 0.750. In comparison, the TS framework improves the transfer performance on the same test device set (Fig. S10c), as shown in the substantially recovered prediction precision (Fig. S10d, NRMSE=0.166), by training the readout on temporally concatenated states from multiple device sets.

The TS framework also shows flexibility in its operation. Besides the working mode I (default unless otherwise stated) shown in Fig. \ref{fig3}a, the TS training can also be operated where the input signals are identical to the different devices (named mode II, see Methods). Mode II also successfully improves the transfer performance for both the single-channel and the three-channel RC (Fig. S11), where the prediction errors on the same test device or device set are only marginally different from those in mode I (testing NRMSE: single channel 0.172; three-channel 0.161). At the same time, the temporal order of the devices/device sets can be permuted. When temporally permuted, the mode II is strictly equivalent to that before, while the mode I shows only slight differences in prediction error (Supplementary Note 2.3, and Fig. S12). Also, the workflow of the TS framework is compatible with the recursive least squares (RLS) algorithm to train the readout online (see Supplementary Note 2.4) \cite{tamura2021transfer, sussillo2009generating}. In addition, as a complementary case study, the TS framework is tested in recurrent Lorenz63 forecasting \cite{gauthier2021next, lorenz1963deterministic, pathak2018model} using the virtual ${\rm TiO}_x$-based RC (see Supplementary Note 1). Each virtual device is generated by a calibrated ${\rm TiO}_x$ dynamical model with experimentally characterized device variations (Figs. S13 and S14). The model is used only to generate reservoir states, whereas TS training still operates on the collected states without using model parameters. Because predictions are recursively fed back as inputs, this task amplifies transfer errors in the classical framework, resulting in failed tracking and attractor reconstruction; in contrast, the same RC in the TS framework successfully tracks the ground-truth for longer time and reconstructs the attractor (Fig.~S15). To briefly summarize, in the benchmarks, the TS framework improves the transfer performance on unseen device/device set in the presence of the D2D variations.

\subsection*{Direct-transfer improvement in classification task}

\begin{figure}[h]
\centering
\includegraphics[width=1\textwidth]{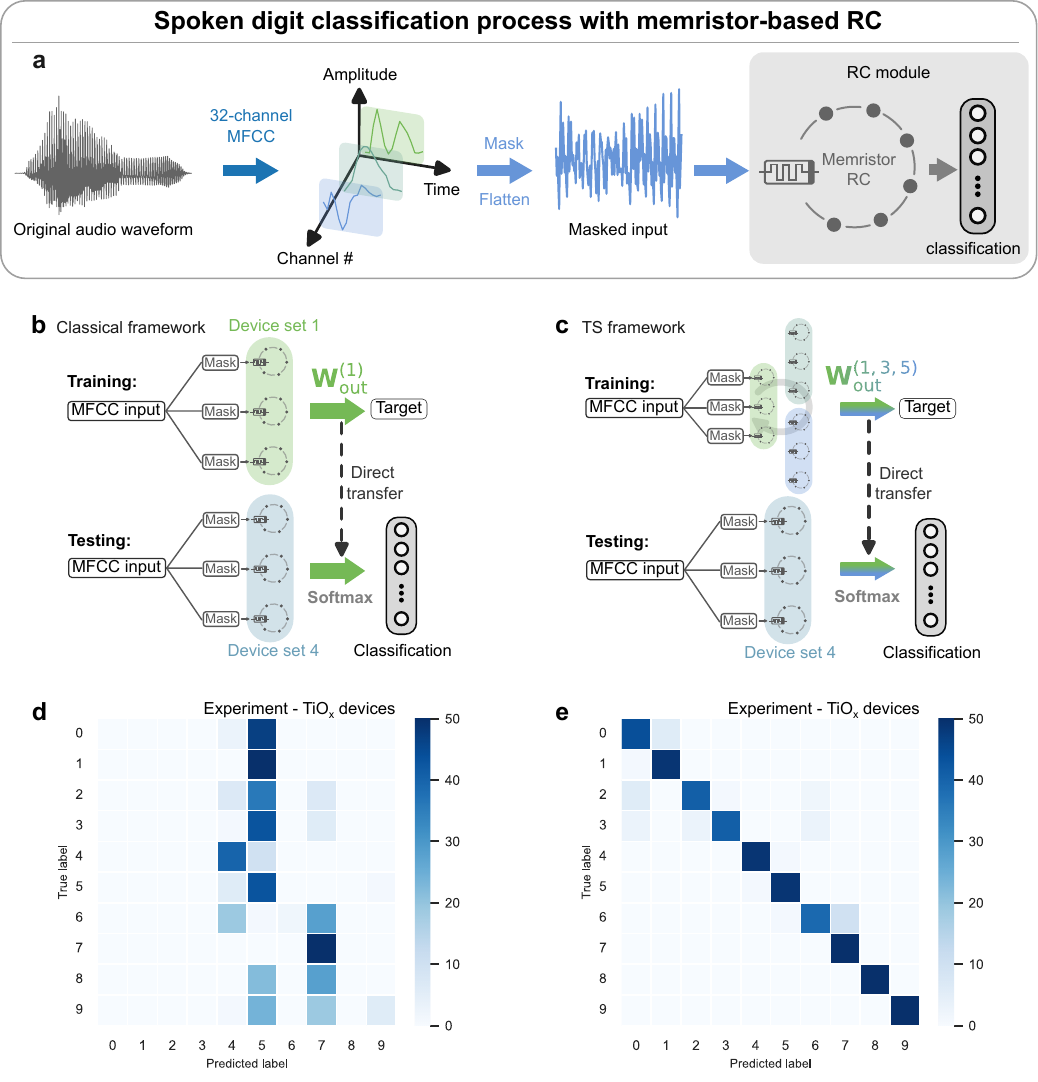}
\caption{Temporal-switch in the spoken digit classification. ({\bf a}) The RC processing of the audio signal of a spoken digit. ({\bf b}) The classical training and transfer of the readout. ({\bf c}) The training and transfer of readout in the TS framework. ({\bf d}) Classification result of the classical framework (accuracy=28.2\%). ({\bf e}) Classification result of the TS framework (accuracy=92.4\%). Both {\bf d} and {\bf e} show the summed testing results of 10-fold cross validation. Here, device sets 1, 3, 4, 5 refer to (9;1;1), (1;9;9), (6;9;1) and (8;9;1), respectively.}\label{fig4}
\end{figure}

We next evaluate the TS framework in the spoken digit classification task \cite{TI46, zhong2021dynamic, caoyi2024physical}. Different from the prediction benchmark, the inputs here are sample-dependent temporal signals with class-specific features, and the transferred readout needs to preserve discriminative information on unseen device/device set. In this task, the original audio waveforms are first converted into 32-channel Mel-frequency cepstral coefficient (MFCC) signals, which are then masked, flattened, and applied to the memristor-based RC. The readout is trained with ten-dimensional one-hot digit label signal and the reservoir state responses. During testing, channel outputs are temporally averaged and classified by a SoftMax-based winner-take-all rule (Fig. \ref{fig4}a, detailed in Methods).

The evaluation is firstly implemented using the three-channel ${\rm TiO}_x$-based RC, by comparing the classification accuracy between the classical (Fig. \ref{fig4}b) and the TS frameworks (Fig. \ref{fig4}c). In the classical framework, the transferred readout causes poor classification compared with the self-training optimum (Fig. S16c), as indicated by the confusion matrices. For example, the readout trained on device set 1 produces the overall accuracy of 28.2\% on device set 4 (Fig. \ref{fig4}d), the one trained on device set 3 produces 67.8\%, and the one trained on device set 5 produces 81.8\% (Figs. S16a,b). While the classification degrades substantially when the readout is trained on any of the three device sets, it is substantially recovered when the readout is trained in the TS framework with the same three device sets. The same test device set (i.e., device set 4) now achieves the accuracy of 92.4\% with the TS-trained readout (Fig. \ref{fig4}e). The confusion matrices, where most spoken digits are assigned to their correct classes (Fig. S17), indicate that the class-discriminative temporal features remain accessible after direct transfer. The improvement in direct transfer is not tied to a specific device family or RC configuration. The TS framework is also tested on the single-channel ${\rm NbO}_x$-based reservoirs, which exhibit different electrical response characteristics. While the ${\rm NbO}_x$-based RC also suffers from the substantially degraded performance in classical direct transfer (Fig. S18sa,b, accuracy=13.8\%), its performance is substantially recovered and close to the self-training optimum (Figs. S18e,f, accuracy=88\%) in the TS framework (Figs. S18c,d, accuracy=89.4\%). In addition, the TS framework is tested in a complementary case study of threshold-based arrhythmia detection \cite{zhong2022memristor, goldberger2000physiobank} (Fig.~S19, detailed in Supplementary Note~1.3). This task also uses the virtual ${\rm TiO}_x$-based RC with experimentally characterized device variations, and the readout training is still based only on generated reservoir states (Fig.~S19a). In the classical framework the network output exhibits clear deviation and causes poor accuracy with the transferred readout and thresholds, whereas in the TS framework the output produces high accuracy, precision and recall rate of arrhythmia detection (Figs. S19c,d).

To briefly summarize, the spoken digit classification task demonstrates that the TS framework successfully improves direct transfer, and the improvement is observed across different device families and network configurations. The complementary arrhythmia detection case study also provides support for TS framework's performance in a threshold-based classification task. Together, these results show that training the readout on multi-device/device set reservoir responses improves classification transfer under the same direct transfer setting.

\subsection*{Performance advantage and robustness}

\begin{figure}[h]
\centering
\includegraphics[width=1\textwidth]{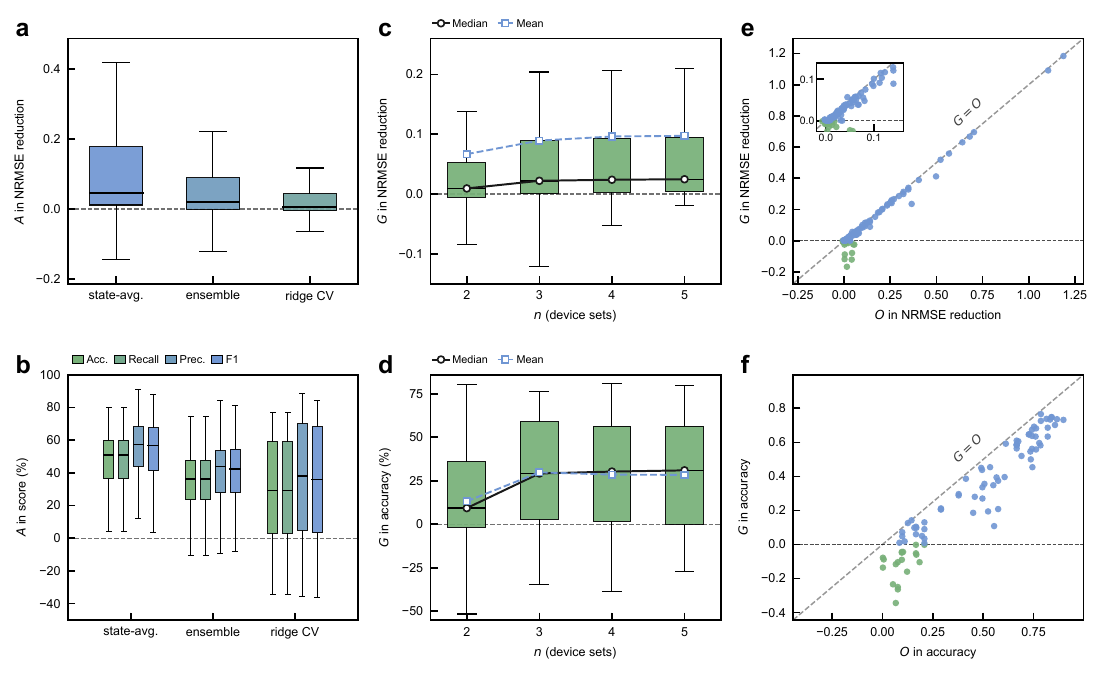}
\caption{Advantage and robustness of temporal-switch. ({\bf a,b}) The TS advantage ($A$) over other baselines in ({\bf a}) the MG benchmark and ({\bf b}) in spoken digit classification ($n=3$). Positive \(A\) indicates that TS outperforms the corresponding baseline. ({\bf c,d}) Influence on the recovered gain ($G$) by the number of training device sets ($n$) in ({\bf c}) the MG benchmark and ({\bf d}) spoken digit classification. Positive $G$ indicates that the TS outperforms the classical direct transfer. ({\bf e,f}) The recovered gain ($G$) from the opportunity ($O$) in ({\bf e}) MG prediction NRMSE and in ({\bf f}) spoken digit classification accuracy ($n=3$; green points: negative gain; blue points: positive gain). {\bf a--f} show trials from all the ``in-range'' cases.}\label{fig5}
\end{figure}

To evaluate the advantage of the TS framework in improving direct transfer performance, we compare it with three baselines using the three-channel ${\rm TiO}_x$-based RC: state-average (state-avg.) training, cross-validated ridge regression (ridge CV), and ensemble method (Fig.~S20; see Methods for details). For a fair comparison, the same training and test device sets are used for the TS framework, the state-average training, and the ensemble method. Ridge CV serves as an optimized single device set transfer baseline, where the training device set is chosen as the first one in the corresponding TS training order. We quantify the pairwise advantage by \(A\), which is defined such that \(A>0\) indicates better direct transfer performance of the TS framework than the corresponding baseline (Methods). In the MG prediction, the TS framework shows advantages in direct transfer, as the \(A\) distributions are shifted above zero and the boxes remain in the positive region for all baseline comparisons (Fig.~\ref{fig5}a). The advantage is larger over state-average training and the ensemble method, both of which also use multiple training device sets, indicating that the improvement does not simply arise from using more training device sets. The advantage becomes smaller but remains positive over ridge CV, suggesting that the TS framework provides benefits beyond optimizing the ridge regularization. A similar trend is observed in spoken digit classification. The boxes of \(A\) for testing accuracy, precision, recall, and \(F_1\) score also remain in the positive region, showing that the TS framework improves direct transfer classification performance over the baselines (Fig.~\ref{fig5}b). The same tendency is observed in the arrhythmia detection case study, where the TS framework better preserves detection performance after direct transfer (Fig.~S19d).

Since the TS framework uses multiple device sets, its performance may be influenced by $n$, the number of device sets used for training. To evaluate this influence, matched training combinations are generated from each \(n=3\) seed combination to ensure fair comparison. One device set is randomly removed to obtain the corresponding \(n=2\) combination, whereas randomly selected device sets are sequentially added to obtain the corresponding \(n=4\) and \(n=5\) combinations. The test device set is always excluded from training. The influence of $n$ on the performance of the TS framework is reflected in two aspects. On one hand, the TS framework retains advantages compared to the other baselines for different $n$. As observed for the default \(n=3\) setting (Figs. \ref{fig5}a,b), the TS framework also shows advantages over the competing baselines when \(n=2\), 4, and 5 in both the MG benchmark and spoken digit classification, as indicated by the positive \(A\) (Fig.~S21). On the other hand, $n$ influences the recovered gain $G$, which is defined such that the positive value indicates improved transfer performance of the TS framework over the classical framework (Methods). Increasing \(n\) from 2 to 3 clearly increases \(G\), with increases in both its mean and median. The increased \(G\) corresponds to lower testing NRMSE in the MG benchmark (Fig. \ref{fig5}c), or higher accuracy (Fig. \ref{fig5}d), precision (Fig. S22a), recall (Fig. S22b) and $F_1$ score (Fig. S22c) in the spoken digit classification. Further increasing \(n\) from 3 to 4 or 5 produces only limited additional improvement. These results indicate that the TS framework benefits from broader measured D2D variation coverage during training, but the gain tends to saturate after a sufficient number of representative training device sets is included.

Although increasing \(n\) improves the recovered gain up to a saturation regime, \(G\) remains broadly distributed across different transfer cases. This dispersion can be understood by considering the available room for improvement in each transfer case. We define the performance opportunity \(O\) as the performance gap between classical direct transfer and the self-training optimum reference (Methods), and \(G\) thus corresponds to the part of this gap recovered by the TS framework. Therefore, \(G=O\) indicates complete recovery to the self-training reference. In the MG benchmark, most transfer cases lie close to the \(G=O\) line, showing that the TS framework recovers nearly all of the available opportunity in these cases (Fig.~\ref{fig5}e). Only a small number of cases near $O=0$ fall below $G=0$, where the classical direct transfer is already near-optimal and the broader transfer optimization can introduce a penalty. In spoken digit classification, the distribution is broader and the points do not lie as close to the \(G=O\) line as in the MG benchmark (Fig.~\ref{fig5}f and Figs.~S22d--f). Nevertheless, most cases still show positive \(G\) with only a modest deviation from the \(G=O\) line, indicating considerable recovery of the available opportunity. A small subset of cases shows positive \(G\) but exhibits noticeable deviations from the \(G=O\) line, corresponding to incomplete recovery. Cases with negative \(G\) account for only a small fraction of the transfer cases, and only a few of them show more pronounced degradation. These results demonstrate the efficacy of TS framework in improving the performance and approaching the self-training optimum for most transfer cases.

A theoretical analysis based on the general RC formulation further clarifies the mechanism behind this improvement (Supplementary Note~2). Unlike the ensemble method, which averages independently optimized device-specific readouts, or the state-average training, which removes device-dependent state dispersion before optimization, the TS framework trains a single readout on the pooled reservoir states formed by temporal-switch concatenation. This optimization jointly incorporates the state--target correlations and state covariances of multiple reservoirs, enabling different reservoir responses to be optimized together under one shared readout and keeping information about D2D variations. The analysis also predicts the relationship between recovered gain and D2D variation strength, which is supported by the ${\rm TiO}_x$-based RC experiments (Fig.~S23).

Additional comparisons further support the advantage. Compared with few-shot calibration approaches \cite{stenning2024neuromorphic}, the TS framework does not require calibration data from the test device set. Notably, it outperforms classical few-shot calibration when only a limited number of target samples are available, including both low- and moderate-shot regimes (Fig.~S24). At the same time, it has a clear performance boundary. Using the experimentally measured \(\Delta I(t_1)\) as a variation indicator (see Supplementary Note 3), test device sets that fall within the training range (``in-range'', see Fig.~S25) show significantly better transfer performance than those outside this range, such as lower NRMSE in MG prediction or higher accuracy, precision and $F_1$ score in spoken-digit classification (Fig.~S26). Together, these results show that the TS framework improves direct transfer performance by effectively sampling D2D variations during training, while its reliable improvement depends on whether the test reservoirs remain within the variation range covered by the training ones.

\section*{Discussion}\label{sec3}

D2D variation remains a major barrier to translating physical computing systems from individually optimized demonstrations to reliable applications \cite{yan2024emerging, liang2024physical, zhong2021dynamic, moon2019temporal}. For physical RC, it directly affects the transferability, which refers to the ability to reuse a trained readout across nominally identical reservoirs. In the classical framework, a readout trained on a single device/device set can incur clear performance degradation when directly transferred to unseen devices, even within the same device family. To address this issue, we introduce the TS framework, which trains one shared readout through the ridge regression over pooled measured states collected from multiple devices/device sets. It achieves reliable direct transfer of the readout, with substantially recovered performance on the test device/device set unseen in the training phase. These results demonstrate the role of the TS framework as an important step towards application-oriented plug-and-play uses.

The effectiveness of the TS framework can be understood from the diversity of measured reservoir states used during training. In the classical framework, a readout is optimized for the state–target correlations and covariance structure generated by one single device/device set, and may therefore become specialized to its state representation. When a deployment device/device set produces shifted states because of the D2D variations, its output processed by the original readout may no longer align well with the target. By contrast, the TS training exposes the shared readout to a broader range of state responses, making the learned mapping less dependent on any specific device/device set. In mode I of the TS framework, different temporal segments of the training input are assigned to different devices, covering both temporal input diversity and device-response variation without increasing the total number of training samples. In mode II, identical inputs are applied to different devices to examine multi-device response coverage more directly. The agreement between the two modes, together with the theoretical analysis, supports the interpretation that the improvement comes from measured-state multi-device coverage. This distinguishes the TS framework from several alternative direct-transfer strategies. Ridge CV improves the regularization but still lacks information about the state diversity caused by D2D variations. The ensemble method averages multiple readouts trained individually on each reservoir, whereas the state-average method averages device states before readout training. Both approaches lose useful information about D2D variations in the covariance structure. Few-shot calibration can adapt to the test device/device set, but requires test device/device set data and therefore does not satisfy the direct transfer setting.

The improvement provided by the TS framework should be interpreted within a transfer range. Its reliability depends on whether the deployment device or device set is sufficiently represented by the variation range covered during the multi-device training. Analysis based on the indicators [e.g., $\Delta I(t_1)$ in this work] provides a practical way to assess this range. Direct transfer is more reliable when the $\Delta I(t_1)$ coordinates of test devices/device sets fall within the range covered by the training devices than when they fall outside it. This transfer range provides a practical design principle for deployment. In application-oriented training, representative devices should be selected to cover the expected variation range in testing/inference. The central objective of this work is to address the readout-transfer degradation caused by D2D variation, which is a key obstacle for reliable application-oriented physical RC. Within this scope, it provides a practical way to improve direct transfer in the presence of D2D variations. Other factors related to deployment, including independent fabrication batches, long-term ageing and environmental drift, may introduce additional sources of variation, and are best addressed through dedicated future studies. The tasks demonstrated in this work cover two broad types of tasks, temporal signal prediction and classification. Therefore, it provides a functional basis for extending the TS framework to more complex and specific real-world applications. Such extensions will require task-specific validation in future studies, but the present results establish the foundation for that direction.

The TS framework is model-free in the sense that the readout training is performed directly on measured states, without requiring an explicit model of device dynamics. This makes the method well suited to physical RC systems where the devices are used as computational black boxes to generate states as responses to external input. At the same time, the TS framework is simple in its operation, which is not a weakness, but a key part of its practical value. It provides an experimentally validated way to improve the degraded direct transfer performance caused by D2D variations, without complicating the inference operation. Integrated implementations will be an important next step, and will require dedicated future efforts in the analysis of complete system-level energy, latency and area. Moreover, because the TS framework is model-free and can be analysed under a general RC formulation, it may be extendable beyond the memristor-based systems validated here to other physical RC platforms, including photonic \cite{aadhi2026scalable, wu2023hybrid}, spintronic \cite{gartside2022reconfigurable, torrejon2017neuromorphic} and mechanical \cite{guo2024MEMS, kartal2025NEMS} reservoirs. By combining model-free training with readout-only deployment, the TS framework provides a practical path towards transferable physical RC hardware for broader application-oriented computing.

\section*{Methods}

\subsection*{Memristor fabrication}

${\rm TiO}_x$ memristor fabrication: The Ti/${\rm TiO}_x$/Pt--based dynamic memristors are fabricated as crossbar structure on a Si wafer with 300-nm-thick thermally grown ${\rm SiO}_2$. First, 10-nm Ti and 30-nm Pt are electron-beam evaporated and patterned as the adhesion layer and bottom electrode, respectively. Then, the functional 10-nm-thick ${\rm TiO}_x$ is deposited by reactive sputtering in the mixed atmosphere of Ar and ${\rm O}_2$ with a ratio of 5:1. Finally, a 30-nm-thick Ti is sputtered and patterned as top electrode. All devices involved in this paper are placed separately, not in a memristor array. The line widths of the electrodes are 5 $\mu$m.
\\\\
\noindent
${\rm NbO}_x$ memristor fabrication: First, a 5/40 nm Ti/Au layer is deposited on the ${\rm SiO}_2$ substrate, followed by lift-off to obtain the patterned bottom electrode. Next, a 20 nm ${\rm Ti}$-doped ${\rm Nb_2O_5}$ layer is prepared using atomic layer deposition by alternately exposing titanium tetrachloride (${\rm TiCl}_4$) and niobium ethoxide (${\rm NbOEt}$) source material, followed by an ${\rm Ar}$ ion-enhanced plasma etching process to expose the bottom electrode. Finally, a 40 nm ${\rm Au}$ layer is deposited by electron beam evaporation. ${\rm Au/Nb_2O_5:Ti/Au}$ devices with crossbar structure are obtained after lift-off and the area of the cross-point is 5 $\mu$m × 5 $\mu$m.

\subsection*{Electrical measurement}

The DC I-V characterization, the pulse programming and decay characterization are all operated on Keysight B1500A Semiconductor Device Analyzer. In the implementation of memristor-based RC, the generation of the original input, the time-multiplexing, training and testing/inference are completed on PC. The application of the voltage signals and the sampling of the electrical currents are completed with the B1500A. For the ${\rm TiO}_x$ memristor, the top electrode is grounded, and the input voltage is applied on the bottom electrode. For the ${\rm NbO}_x$ memristor, the bottom electrode is grounded, with the input applied on the top electrode.

\subsection*{Characterization}

\subsubsection*{I--V characteristics}

For ${\rm TiO}_x$, the I--V characteristics is characterized by sweeping the applied voltage from 0 V to 2.5 V, then decreasing it to -1 V, and finally returning it to 0 V. For ${\rm NbO}_x$, the I--V characteristics is characterized by sweeping the applied voltage from 0 V to 4 V, then returning it to 0 V.

\subsubsection*{Pulse programming}

Five consecutive ‘write’ pulses are applied on each memristor and the electrical current is recorded as response. The pulse amplitude is 3 V for ${\rm TiO}_x$, and 2.5 V for ${\rm NbO}_x$. The voltage in the interval is 0.5 V. The pulse width is 200 ms and the interval is 300 ms. $\Delta I(t_1)$ is calculated as the difference between the electrical currents sampled at the first and the last moment of the first pulse. $\Delta I(t_2)$ is calculated as the difference in the second pulse. In this setting, $\Delta I(t_1)$ is the increment of currents under the same pulse (the same amplitude and width), therefore corresponding to the change rate in conductance, i.e., the dynamic response to external voltage stimulus.

\subsubsection*{Decay effect}

A `write' pulse (3 V, 1 ms) is firstly applied to each device, followed by a series of short `read' pulses (pulse: 2 V, 10 $\mu$s; interval: 0.5 V, 10 $\mu$s) to measure and record the electrical currents at given moments (e.g., at the end of every `read' pulse) during the decay process (see Figs. S6a,b). The decay time constant $\tau$ is calculated by fitting the recorded data to an exponential decay relation by least squares method (Fig. S6c) as in
\[
    i(\Delta t) = b_1 + \exp\left(-\frac{\Delta t}{\tilde{\tau}} + b_2\right),
\]
where $b_1, b_2$ are parameters to be automatically fitted by the algorithm and $\Delta t$ is the time since the end of `write' operation. $\tilde{\tau}$ is the fitted value of $\tau$. We use $\tau$ to represent $\tilde{\tau}$ elsewhere without causing ambiguity.

\subsection*{Memristor-based RC}

All memristor-based RCs involved in this work are implemented in the single node structure \cite{zhong2021dynamic}, which is equivalent to the classical RC structure but requires far fewer devices for implementation \cite{appeltant2011information}. The network is shown in Fig. S1. It consists of three steps in data processing: the time-multiplexing, the sampling of electrical current and the conversion to high-dimensional reservoir states.

\subsubsection*{Time-multiplexing}

The original signal ${\bf x}_t$ [discrete, or sampled from the continuous signal ${\bf x}(t)$] is processed into a one-dimensional step-like continuous signal. Firstly, the original input ${\bf I}_{\rm in}=\{{\bf x}_t\}$ (in the form of $L\times T$, $L$ is the dimension of original input, $T$ is the time length) is multiplied by a binary input weight matrix ${\bf W}_{\rm in}$ (in the form of $N\times L$, $N$ is the size of reservoir, i.e., the number of virtual nodes, the absolute value of its elements is $\beta$) to obtain the high-dimensional intermediate input ${\bf J} = {\bf W}_{\rm in}{\bf I}_{\rm in}$ (in the form of $N\times T$). Secondly, ${\bf J}$ is flattened into a one-dimensional discrete signal, which is then transformed into a continuous step-like signal ${J(t)}$ through a `sample \& hold' operation with the time unit of $\theta$ (Fig. S2). $J(t)$ is finally re-scaled to a given voltage range of $[v_1, v_2]$ through linear mapping to get the input voltage signal $v(t)$ to memristor as
\[v(t) = \frac{[J(t) - (a+b)/2]}{a-b}\times(v_2 - v_1) + \frac{v_1+v_2}{2},\]
where $a = {\rm max}[J(t)], b = {\rm min}[J(t)], t\in [0, N\cdot T\cdot \theta]$. The set $(v_1, v_2, a, b)$ is named scaling factor. The same scaling factor indicates the same re-scaling process of $J(t)$ to $v(t)$. Also, the same re-scaling process and the same input weight matrix ${\bf W}_{\rm in}$ together indicate the same time-multiplexing. 

\subsubsection*{Sampling electrical current}

The internal states (e.g., conductance) of the memristor changes when  stimulated by the external voltage $v(t)$, which results in the electrical current $i(t)$ as response. The values of $i(t)$ at the end of each $\theta$ are sampled and collected as one-dimensional discrete electrical current signal $i(k)$ where $k=1, 2, ..., N\cdot T$ (Fig. S3). 

\subsubsection*{Conversion to reservoir states}

The sampling of electrical current $i(k)$ at a series of given moments is the process of reading the states of the `virtual nodes'. The high-dimensional reservoir states of one iteration (named frame in the following) are given by $N$ consecutive sampled currents. Within each frame, the virtual nodes are placed with the distance of $\theta$ to the adjacent nodes. The state series is split into different frames ${\bf r}_1^\top,{\bf r}_2^\top,{\bf r}_3^\top, ..., {\bf r}_T^\top$, which are then concatenated as high-dimensional states matrix ${\bf R} = \left[{\bf r}_1, {\bf r}_2, {\bf r}_3, ..., {\bf r}_T\right]$ (Fig. S3). Then, the readout of the reservoir is trained as 
\begin{equation}\label{ridge regression}
    {\bf W}_{\rm out} = {\bf Y}{\bf R}^\top\left({\bf R}{\bf R}^\top+\lambda{\bf I}\right)^{-1},
\end{equation}
where ${\bf Y}$ is the target (in the form of $M\times T$, $M$ is the dimension of output), ${\bf I}$ is the identity matrix and $\lambda$ is the ridge parameter.

\subsubsection*{Multi-channel configuration}

In the multi-channel parallel configuration, several devices operate simultaneously to provide an enlarged reservoir representation (i.e., increased dimension). Each channel receives the same input signal through its assigned input mask and generates its own reservoir states as response independently. For a configuration containing $N_{\rm mc}$ channels, the reservoir-state vector at frame $t$ is jointly constructed by the states from all channels:
\[
{\bf r}_t=\left[{\bf r}_{1, t}^\top,{\bf r}_{2, t}^\top,\ldots,{\bf r}_{N_{\rm mc}, t}^\top\right]^\top ,
\]
where ${\bf r}_{i,t}$ denotes the state vector of the $i$-th channel. The resulting state matrix is then used to train a single readout by ridge regression, following the same procedure as in \eqref{ridge regression}.  The input masks and time-multiplexing settings are kept fixed for the corresponding channels during training and testing. The device set is written as $(d_1;d_2;...;d_{N_{\rm mc}})$, where $d_1,~d_2,...,d_{N_{\rm mc}}$ are the devices used in channel $1,~2,~...,~N_{\rm mc}$, respectively. It is ordered because the channels are treated as different physical channels.

\subsection*{TS workflow}

The workflow diagram of the TS framework is shown in Fig. S8. The workflow is divided into two phases: the training process and the testing process.
\\\\
\noindent
Training: Start from $k=1$. Input the signal $\{{\bf x}_{(k),t_k}\} = {\bf X}_{(k)} = [{\bf x}_{(k),1}, {\bf x}_{(k),2}, ..., {\bf x}_{(k),T}]$ (in the form of $L\times T$) to the device/device set $k$, run the device-based RC (including preprocessing), and collect the response reservoir states $\{{\bf r}_{(k),t_k}\}={\bf R}_{(k)} = [{\bf r}_{(k),1}, {\bf r}_{(k),2}, ..., {\bf r}_{(k),T}]$ (in the form of $N\times T$). The corresponding target signal is $\{{\bf y}_{(k),t_k}\} = {\bf Y}_{(k)} = [{\bf y}_{(k),1}, {\bf y}_{(k),2}, ..., {\bf y}_{(k),T}]$ (in the form of $M\times T$). The states and target signal are concatenated sequentially with the existing record of states and target:
${\bf R}_{(1,...,k)} = [{\bf R}_{(1,...,k-1)},{\bf R}_{(k)}],~{\bf Y}_{(1,...,k)} = [{\bf Y}_{(1,...,k-1)},{\bf Y}_{(k)}]$. Then $k=k+1$. If $k\leq n$, repeat the signal input, state collection and concatenation above. If $k > n$, train the readout
\begin{equation}\label{TS regression}
   {\bf W}_{\rm out}^{(1,2,...,n)} = {\bf Y}_{(1,...,n)}{\bf R}_{(1,...,n)}^\top\left({\bf R}_{(1,...,n)}{\bf R}_{(1,...,n)}^\top+\lambda{\bf I}\right)^{-1}. 
\end{equation}
Then the training phase finishes. $n$ is the number of training devices/device sets used in the TS framework. For the working mode I shown in Fig.~\ref{fig3}a, the input to each device is different, i.e., $\{{\bf x}_{(i),t_i}\}\neq\{{\bf x}_{(j),t_j}\}$ for $i\neq j$. For the working mode II shown in Fig. S11, the input to each device is identical, i.e., $\{{\bf x}_{(i),t_i}\}=\{{\bf x}_{(j),t_j}\}$ for $i\neq j$.
\\\\
\noindent
Testing: Select the test device. Input the signal $\{{\bf x}_{({\rm ts}),t_{\rm ts}}\} = {\bf X}_{({\rm ts})} = [{\bf x}_{({\rm ts}),1}, {\bf x}_{({\rm ts}),2}, ..., {\bf x}_{({\rm ts}),T_{\rm ts}}]$ to the test device, run the RC and collect the response states $\{{\bf r}_{({\rm ts}),t_{\rm ts}}\}={\bf R}_{({\rm ts})} = [{\bf r}_{({\rm ts}),1}, {\bf r}_{({\rm ts}),2}, ..., {\bf r}_{({\rm ts}),T_{{\rm ts}}}]$. The prediction in the testing/inference phase is calculated as ${\hat{\bf Y}} = {\bf W}_{\rm out}{\bf R}_{({\rm ts})}$. Here, $T_{({\rm ts})}$ is the time length for the test device/device set.

\subsection*{Baselines for comparison}

In this work, the TS framework is compared to three direct-transfer baselines. Like in the TS framework, the readout is trained before deployment and is directly transferred to unseen device or device set without any post-training calibration or adjustment in all these three baselines. Unless otherwise stated, the reservoir configuration, the time-multiplexing, the data and the dimensionality are all identical to those in the TS framework.

\subsubsection*{Ensemble}

The ensemble method trains multiple single-device (or single-device set) readouts independently (Fig. S20a). For the $n$ selected devices (or device sets) used for training, the reservoir state matrix ${\bf R}_{(i)}$ from device/device set $i$ is used to train a device or device set-specific readout ${\bf W}_{\rm out}^{(i)}$, with the same ridge-regression procedure, and the same training input and target signals.

During testing, the reservoir states of the unseen device or device set are processed by the average of all trained readouts $\bar{\bf W}_{\rm out}=\frac{1}{n}\sum_{i=1}^n{\bf W}_{\rm out}^{(i)}$. The final prediction is obtained as $\hat{\bf Y}_{\rm ens}=\bar{\bf W}_{\rm out}{\bf R}_{({\rm ts})}$. For classification tasks, the averaging is performed on the output scores before the final decision rule. 

To clarify, the ensemble baseline is related to multi-channel parallel RC because both use multiple device responses, but they differ critically in their objectives and settings. Multi-channel parallel RC is an architectural expansion where multiple reservoir channels operate simultaneously and their states are concatenated, not temporally, but spatially, to train one readout, thereby increasing the reservoir representation. The ensemble method, instead, trains multiple single-device readouts independently and averages them at testing (and thus, does not increase dimension). It is therefore not an expanded architecture, but a control for testing whether the transfer improvement can be explained by combining several device-specific readouts.

\subsubsection*{State--average}

The state--average (state-avg.) method uses multiple training devices but averages their reservoir responses before readout training (Fig. S20b). For aligned reservoir-state matrices (${{\bf R}_{(i)}},~i=1,2,...,n$), which are obtained from the selected training devices as response to the same input, the averaged state matrix is computed as
\[
\bar{\bf R}=\frac{1}{n}\sum_{i=1}^{n}{\bf R}_{(i)}.
\]
A single readout is then trained by ridge regression using $\bar{\bf R}$ and the corresponding target ${\bf Y}$, and is directly transferred to the unseen test device/device set. This baseline preserves the single-readout deployment form but removes device-dependent state dispersion for training. It therefore tests whether the multi-device responses can be compressed into an average reservoir representation without losing transferability.

\subsubsection*{Ridge CV}

The cross-validated ridge regression (ridge CV) follows the classical  single-device/device set training procedure (Fig. S20c), but selects the optimal readout regularization strength by cross-validation. The ridge parameter $\lambda$ is chosen from a predefined candidate set $\Lambda=\{0,10^{-16},10^{-14},10^{-12},10^{-10},10^{-8},10^{-6},10^{-4},10^{-2},1,10^{2}\}$. For each candidate value, the training data are split into five contiguous folds. A readout is trained on four folds and evaluated on the held-out fold, and the validation errors are averaged over the five folds. The validation error is computed as the NRMSE between the readout output and the target signal (for both the MG benchmark and the spoken digit classification). The $\lambda$ giving the lowest mean validation NRMSE is then used to retrain the readout on the full  training data. The resulting readout is directly applied to the unseen test device/device set without any calibration or adjustment. Thus, ridge CV tests whether transfer degradation can be mitigated by tuning the readout regularization strength within the classical training framework alone.

\subsection*{MG benchmark}

In the MG prediction ($L=M=1$), the memristor-based RC ($\lambda=0,~N=10,~\theta=20 ~\mu$s) predicts the state of the Mackey--Glass system \cite{mackey1977oscillation}. It is described by a one-dimensional delay differential equation:
\begin{equation}
     \dot x = -b \cdot x(t) + \frac{a \cdot x(t-\tau_{\rm MG})}{1 +x(t-\tau_{\rm MG})^c}.
\end{equation}
The discrete signal based on the MG system is generated as
\begin{equation}
    x(k+1) = x(k) + \Delta t \cdot \left[-b \cdot x(k) + \frac{a \cdot x(k-\tau_{\rm MG}/\Delta t)}{1 + x(k-\tau_{\rm MG}/\Delta t) ^ c}\right],
\end{equation}
where $\tau_{\rm MG}/\Delta t$ is set as an integer. $x(k)=0.01$ for all $k<0$ (initial conditions) to avoid fixed zero state. The MG discrete signal is chaotic with the parameters $a=0.2, b=0.1, c=10, \Delta t=1, \tau_{\rm MG}=18$.

In the task, the original signal $\{{x}_t\}=[x(1), x(2),...x(T)]$ is split into 18 segments, with the corresponding target ${y}_t={x}_{t+1}$. The time length of each segment is 100 frames, with the adjacent ones overlap each other by 20 frames, resulting in an effective time length of $80$ frames. The segments are then processed by the time-multiplexing and rescaled to [2 V, 2.5 V], into voltage signal $v(t)$. The readout is trained as in \eqref{ridge regression} [\eqref{TS regression} for TS]. In both training and testing phases, $T=720$. For multi-channel parallel RC, $N=10\times N_{\rm mc}$, $N_{\rm mc}$ is the number of multiple channels. In this task, the predicted value $\hat{y}_t$ is compared with ground-truth target $y_t$, and the squared error $e_t=(\hat{y}_t-y_t)^2$ is also calculated, with the overall NRMSE reported at the same time.

\subsection*{Spoken digit classification}

In this task, the memristor-based RC classifies the spoken digits into the corresponding classes. The original audio waveforms (from the TI-46 database \cite{TI46}) are firstly resampled at 8000 Hz, then transformed into 32-channel MFCC signals as input $\{{\bf x}_t\}$. The targets $\{{\bf y}_t\}$ are ten-dimensional one-hot encoded signals, where the value of the channel corresponding to the digit class is 1, with the values of all other channels set to 0 ($L=32,~ N=40,~ \theta=20~\mu{\rm s},~M=10,~\lambda=10^{-16}$). Following the protocol previously reported in ref. \cite{zhong2022memristor}, 10-fold cross-validation is performed at the utterance level, stratified by digit class. For each digit (0–-9), 50 utterances are randomly permuted within the class, and split into 10 folds of 5 utterances. The digit samples are presented to the RC in an isolated manner. The MFCC signals are processed by the time-multiplexing and rescaled to [1.8 V, 2.3 V], into the input voltage signals $v(t)$. The reservoir states $\{{\bf r}_t\}$ are obtained after the sampling and conversion. The readout is trained as in \eqref{ridge regression} [\eqref{TS regression} for TS].

In the testing process, the RC generates a ten-dimensional prediction signal $\hat {\bf Y}={\bf W}_{\rm out}{\bf R}_{\rm (ts)}$. The result is averaged within each channel to obtain a ten-dimensional vector. The vector is processed with the SoftMax function to obtain the predicted probability of the classes. The class of the largest probability is then output as the classification result. For the classical framework, $n=1$, the training dataset consists of 450 digits. For the TS computational framework, $n=3$, each training device/device set processes 150 digits. For multi-channel parallel RC, $N=40\times N_{\rm mc}$, $N_{\rm mc}$ is the number of multiple channels. In this task, the classified results are shown in four metrics: accuracy, precision, recall and $F_1$ score.

\subsection*{ANOVA and ICC}

The analysis of variance (ANOVA) is used to evaluate the statistical significance of D2D and C2C variations. It is a statistical technique that compares the means of multiple groups by partitioning the total variability into between-group (between-device, i.e., device-to-device) and within-group (within-a-device, i.e., cycle-to-cycle) components. For a given quantity $x\in\{\Delta I (t_1), \Delta I (t_2),\tau\}$, the data consists of $k$ different groups, each corresponding to one device. The $j$-th value in the $i$-th group is denoted as $x_{ij}$, where $i=1,2,...k$ and $j=1,2,...n_i$. The mean of the $i$-th group is calculated as 
$$
\bar{x}_i = \frac{1}{n_i}\sum_{j=1}^{n_i}x_{ij},
$$
with the overall mean as
$$
\bar{x} = \frac{1}{N}\sum_{i=1}^{k}\sum_{j=1}^{n_i}x_{ij},
$$
where $N=\sum_{i=1}^kn_i$ is the total count of data values. The total variation in data is calculated as the total sum of squares
$$
SS_T = \sum_{i=1}^{k}\sum_{j=1}^{n_i}(x_{ij}-\bar{x})^2.
$$
It is further decomposed into the between-group and within-group components:
$$
SS_T=SS_B+SS_W,
$$
where the between-group sum is 
$$
SS_B = \sum_{i=1}^k n_i(\bar{x}_i - \bar{x})^2,
$$
and the within-group sum is 
$$
SS_W = \sum_{i=1}^{k}\sum_{j=1}^{n_i}(x_{ij}-\bar{x}_i)^2.
$$

The degrees of freedom for the between-group, within-group, and the total variations are respectively
\begin{equation*}
\begin{aligned}
    df_B&=k-1,\\
    df_W & = N-k,\\
    df_T & = N-1.
\end{aligned}
\end{equation*}
The corresponding mean squares are obtained as 
\begin{equation*}
\begin{aligned}
    MS_B&=\frac{SS_B}{df_B}=\frac{SS_B}{k-1},\\
    MS_W & = \frac{SS_W}{df_W}=\frac{SS_W}{N-k},\\
\end{aligned}
\end{equation*}
the observed ANOVA $F$-statistic is then calculated as 
$$
F_{\rm \small obs}=\frac{MS_B}{MS_W}.
$$
The null hypothesis of ANOVA is that all the group means are equal:
$$
H_0 :\mu_1=\mu_2=\ldots=\mu_k;
$$
and the alternative hypothesis is that at least one group mean is different:
$$
H_1 : \exists i,j\quad {\rm such ~that}\quad \mu_i\neq\mu_j.
$$
Under the null hypothesis, the calculated $F$-statistic follows an $F$-distribution with $k-1$ and $N-k$ degrees of freedom: $F\sim F_{k-1, N-k}$. The corresponding $p$-value is obtained from the upper tail of the $F$-distribution: $p=P(F_{k-1, N-k}\geq F_{\rm \small obs})$. For a given significance level of $\alpha=10^{-4}$, when $p<\alpha$, the null hypothesis is rejected, indicating that the observed D2D variations are statistically significant.

In addition to ANOVA, the intra-class correlation coefficient (ICC) is calculated to quantify the proportion of total variations attributable to D2D variations. The measured/calculated quantity is modelled as such: $x_{ij}=\mu+d_i+\epsilon_{ij}$, where $\mu$ is the overall mean. $d_i$ is the device-specific random effect (i.e., D2D variation) and $\epsilon_{ij}$ is the within-device random effect (i.e., C2C variation). They are assumed to follow $d_i\sim N(0, \sigma_{\rm \small D2D}^2),~\epsilon_{ij}\sim N(0, \sigma_{\rm \small C2C}^2)$. Therefore, the total variance is expressed as $\sigma_{\rm \small total}^2 = \sigma_{\rm \small D2D}^2+\sigma_{\rm \small C2C}^2$, and the ICC defined as 
$$
{\rm ICC} = \frac{\sigma_{\rm \small D2D}^2}{\sigma_{\rm \small D2D}^2+\sigma_{\rm \small C2C}^2}.
$$
The variance components can be estimated from the ANOVA mean squares as 
$$
\hat\sigma_{\rm \small C2C}^2=MS_W,~\hat\sigma_{\rm \small D2D}^2=\frac{MS_B-MS_W}{n}.
$$
Therefore, the ICC can be calculated as 
$$
{\rm ICC}=\frac{MS_B-MS_W}{MS_B+(n-1)MS_W}.
$$

\subsection*{NRMSE}

The normalized root mean squared error (NRMSE) is introduced to evaluate the precision of RC in the MG benchmark. Let ${\bf y} = [y(1), y(2), ..., y(T)]$ be the ground truth and ${\hat{\bf y}} = [\hat{y}(1), \hat{y}(2), ..., \hat{y}(T)]$ be the prediction generated by RC. The NRMSE is calculated as  $${\rm NRMSE}({\bf y}, {\hat{\bf y}}) = \sqrt{\frac{1}{T} \cdot \frac{\sum_{i=1}^{T}[{\hat{y}(i)}-{y}(i)]^2}{{\rm var}({\bf y})}},$$ where $T$ is the time length and ${\rm var}({\bf y})$ is the variance of ${\bf y}$. Lower NRMSE value indicates better precision. In Fig.~S23a, the normalized mean squared error (NMSE) is also used to evaluate prediction error, where
\begin{equation*}
\begin{aligned}
{\rm NMSE}({\bf y}, {\hat{\bf y}}) & = \frac{1}{T} \cdot \frac{\sum_{i=1}^{T}[{\hat{y}(i)}-{y}(i)]^2}{{\rm var}({\bf y})}\\
& = \left[{\rm NRMSE}({\bf y}, {\hat{\bf y}})\right]^{2}.\\
\end{aligned}
\end{equation*}

\subsection*{Accuracy, recall, precision \& $F_1$ score}

In spoken digit classification, the performance is evaluated using accuracy, precision, recall, and $F_1$ score. The predictions are compared with the ground-truth labels to construct a multi-class confusion matrix $(C_{i,j})_{10\times10}$. Here, $C_{i,j}$ denotes the number of samples whose true class is $i$ and predicted as $j$. The numbers of true positives (TP), false positives (FP), and false negatives (FN) are calculated for each true class $c$ as follows:
$$
{\rm TP}_c = C_{c,c}, ~{\rm FP}_c = \sum_{i\neq c} C_{i,c},~{\rm FN}_c=\sum_{j\neq c}C_{c,j}.
$$
Accuracy (acc.) measures the fraction of correctly classified samples among all samples:
\[
{\rm Accuracy}=\frac{\sum_cC_{c,c}}{N_{\rm tot}},
\]
where $N_{\rm tot} = \sum_{i,j}C_{i,j}$ is the total number of samples.

Precision (prec.), recall and $F_1$ score ($F_1$) are evaluated in a one-versus-rest manner for each digit class. Precision measures the fraction of predicted positive samples that are correctly classified. Recall measures the fraction of positive samples that are correctly detected. The $F_1$ score is the harmonic mean of precision and recall. For class $c$,
\begin{equation*}
\begin{aligned}
{\rm Precision}_c & = \frac{{\rm TP}_c}{{\rm TP}_c+{\rm FP}_c},\\
{\rm Recall}_c&=\frac{{\rm TP}_c}{{\rm TP}_c+{\rm FN}_c},\\
F_{1, c}&=\frac{2\cdot{\rm Precision}_c\cdot{\rm Recall}_c}{{\rm Precision}_c+{\rm Recall}_c}.
\end{aligned}
\end{equation*}
The reported precision, recall and $F_{1}$ score are averaged over all digit classes:
$$
{\rm Precision}=\frac{1}{10}\sum_{c=1}^{10}{\rm Precision}_c,\quad
{\rm Recall}=\frac{1}{10}\sum_{c=1}^{10}{\rm Recall}_c,\quad
F_{1}=\frac{1}{10}\sum_{c=1}^{10}F_{1, c}.
$$
Specifically, for a balanced ten-class test set, each class contains
$N_{\rm tot}/10$ samples. Therefore,
\begin{equation*}
\begin{aligned}
{\rm Recall}
&=\frac{1}{10}\sum_{c=1}^{10}
\frac{{\rm TP}_c}{{\rm TP}_c+{\rm FN}_c} \\
&=\frac{1}{10}\sum_{c=1}^{10}
\frac{{\rm TP}_c}{N_{\rm tot}/10} \\
&=\frac{\sum_{c=1}^{10}{\rm TP}_c}{N_{\rm tot}} \\
&=\frac{\sum_{c=1}^{10}C_{c,c}}{N_{\rm tot}}\\
&={\rm Accuracy}.
\end{aligned}
\end{equation*}
Thus, the averaged recall is equivalent to accuracy when the test set is balanced in classes.

\subsection*{Opportunity, gain \& advantage}

To evaluate the efficacy of the TS framework in recovering RC performance, we define three quantities on target device set $j$. The opportunity $O$ is defined as the performance gap of the classical framework (trained with the first device set in the TS combination) to per-device set optimum:
$$
O = f_{\rm OP}(j) - f_{\rm CL}(j);
$$
the recovered gain $G$ is defined as the performance improvement of the TS framework compared to the classical framework:
$$
G = f_{\rm TS}(j) - f_{\rm CL}(j);
$$
and the advantage $A$ is similarly defined as the performance difference over the other three baseline methods by the TS framework:
$$
A_{\rm b} = f_{\rm TS}(j) - f_{\rm b}(j);
$$
where $f_{\rm  OP}(j)$, $f_{\rm  CL}(j)$ and $f_{\rm TS}(j)$ are the testing performance metrics by device set $j$ with self-training, classical training and TS training. $f_{\rm b}(j)$ corresponds to the baseline method using the matched training device sets as those used in the TS framework, ${\rm b}\in\{\text{ensemble, state-avg., ridge CV}\}$. In the MG benchmark, the metric $f$ is $(-{\rm NRMSE})$; in spoken digit classification, the metric can be accuracy, recall, precision and $F_{1}$ score.

\section*{Declarations}

\bmhead{Data availability}

All data supporting the results and conclusions of this study are available from the corresponding author upon reasonable request. Source data are provided with this paper.

\bmhead{Code availability}

The code used to generate the results of this study is publicly available from \href{https://github.com/dmclatfdu/Codes-for-TS-framework-for-transferable-neuromorphic-computer}{https://github.com/dmclatfdu/Codes-for-TS-framework-for-transferable-neuromorphic-computer}.

\bmhead{Supplementary information}

A Supplementary Information document is attached, consisting of Supplementary Notes 1-3, Supplementary Figures S1-S26 and Supplementary Tables S1-S11.

\bmhead{Acknowledgements}

B.-W.Q. is supported by the National Natural Science Foundation of China (NSFC) (No. 12371482 and No. 12522123). X.Z. is supported by the NSFC (No. 62488101 and No. 62374040), by the STCSM (No. 25LN3201200 and No. 24QA2700400). W.L. is supported by the NSFC (No. 11925103 and No. 12531018), by the STCSM (No. 2021SHZDZX0103, No. 22JC1402500, No. 22JC1401402 and No. 25JS2810400), and by the SMEC (No. 2023ZKZD04 and No. 2023KEJI05-72). Q.L. is supported by the NSFC (No. T2293732).

\bmhead{Author contribution}

B.-W.Q., X.Z., W.L. and Q.L. conceived the idea. C.L., P.C. and X.Z. fabricated the devices. Z.Z., C.L. and P.C. contributed to the electrical characterization and physical implementation of RC. Z.Z., S.C. and B.-W.Q. built the simulated RC and analysed the mechanism. X.Z., B.-W.Q., W.L. and Q.L. supervised the project. All authors wrote the manuscript.

\bmhead{Declaration of interests}

The authors declare no competing interests.

\bibliography{TS_NatCommun}

\end{document}